\documentstyle[12pt]{article}
\def\({\left (}
\def\){\right )}
\def\[{\left [}
\def\]{\right ]}
\def\cR{\hat{{\cal R}}}
\def\aR#1#2{\hat{{\cal R}}_{#1\; #2}}
\def\R#1#2{\hat{{\cal R}}_{[#1][#2]}}

\def\P#1#2{{\cal P}_{[#1][#2]}}
\def\hBR{\hat{{\cal R}}}
\def\A{{\cal A}}
\def\hR{\hat{R}}

\def\be{\begin{equation}}
\def\ee{\end{equation}}
\begin{document}
 \begin{titlepage}
 \vspace*{-4ex}
 \null \hfill MPI-Ph/92-17\\
 \null \hfill February, 1992 \\
\vskip 3 true cm
 \begin{center}
 {\bf \LARGE Classical Isomorphisms for Quantum Groups }\\[14ex]
{\large
  Vidyut Jain and Oleg Ogievetsky }
 \footnote{On leave of absence from P.N.Lebedev Physical
 Institute, Theoretical Department, 117924 Moscow, Leninsky
 prospect 53, CIS.}
   \\ [6ex]
%

Max-Planck-Institut f\"{u}r Physik und Astrophysik \\
  - Werner-Heisenberg-Institut f\"{u}r Physik \\
 P.O.\,Box 40 12 12 , D - 8000 Munich 40, Germany \\ [2ex]
 \end{center}
\vskip 4cm
 \begin{abstract}
 The expressions for the $\hat{R}$--matrices for the quantum groups
 SO$_{q^2}$(5) and SO$_q$(6) in terms of the $\hat{R}$--matrices for
 Sp$_q$(2) and SL$_q$(4) are found, and the local isomorphisms
of the corresponding quantum groups are established.
 \end{abstract}
 \end{titlepage}

\noindent {\Large \bf 1. Introduction.}

Classical \ isomorphisms \ are \ isomorphisms \ between \
low \ dimensional \
groups \ of \ different \ series. \ The \ list \ of \ classical \
isomorphisms \ includes\ \
 SL(2)$\;\sim\;$SO(3)$\;\sim\;$Sp(1), \ \
SL(2)$\;\times\;$SL(2)$\;\sim\;$SO(4),\ \
 Sp(2)$\;\sim\;$SO(5) \  and \ SL(4)$\;\sim\;$SO(6).

All the orthogonal versions of these
groups appear in four dimensional physics. They are:
the group of spatial rotations, the Lorentz group,
the group of transformations
of compactified Minkowski space and the conformal
group. The classical
isomorphisms provide information about the spinor representations of
these groups.

Currently there is much interest in building a realistic physical
model based on quantum groups. Therefore, one needs to explore
$q$--generalizations of these isomorphisms. The $q$--Lorentz group
and $q$--Minkowski space were studied in [1--5]. This concerns the
isomorphism SL(2)$\times$SL(2)$\sim$SO(4). The time coordinate in
the $q$--Minkowski space is central.
One can therefore consistently set
it to zero, which provides a $q$--generalization of
the isomorphism
SL(2)$\sim$SO(3). The existence of the $q$--deformed
$\epsilon$--tensor
for SL(2) explains the isomorphism SL(2)$\sim$Sp(1).

The aim of the present article is to establish $q$--generalizations of
the two last isomorphisms,
\be
{\rm Sp(2) \sim SO(5) \;\; and\;\; SL(4) \sim SO(6).}
\ee

Quantum groups are conveniently described by their coaction upon
``$q$--deformed'' vector spaces [6]. In describing the
$q$--Lorentz group one builds the $q$--Minkowski space out of two
copies of the two dimensional $q$--space for SL$_q$(2). We follow
the same strategy. The $q$--vector spaces for Sp$_q$(2) and
SL$_q$(4) are well known [7].
To obtain the isomorphisms (1) on the classical level,
one considers the action of SL(4) on the antisymmetric tensor
product of two vectors, and the action of Sp(2) on the traceless
part of the antisymmetric product.
For quantum spaces this is not enough, and one has to find
the commutation relations
preserved by this action. There are in general
several possibilities for such commutation relations and
each of them provides an $\hR$-matrix for bivectors.
The one we choose is different from the $q$--Minkowski case.
Our choice of the $\hR$-matrix is motivated by its compatibility
with the projector onto the
antisymmetric tensor product of two vectors.
 This is discussed in Section 2.
In Section 3
we demonstrate the coincidence of the resulting $\hR$-matrix
in the SL$_q$(4) case with the $\hR$-matrix
of SO$_q$(6). In Section 4 the
isomorphism of Sp$_q$(2)
and SO$_{q^2}$(5) is established. Note that unlike the previous case,
here we obtain SO$_{q^2}$ and not SO$_q$.
The check of automorphisms involves
a direct comparison of $\hR$--matrices; we did not discover
elegant arguments to deduce the isomorphisms.

The general comparison of different $\hR$-matrices
in higher dimensions is left
to a future publication. In addition, we will discuss elsewhere the
real forms of the isomorphisms above.

\newpage
\noindent{\Large\bf 2. $\hR$-matrix for bivectors}

First we briefly recall the construction of
classical local isomorphisms Sp(2)$\sim$SO(5) and SL(4)$\sim$SO(6).
Denote by $V_{\rm Sp(2)}$, $V_{\rm SL(4)}$
the defining representations
of the corresponding groups. In both cases the action preserves the
tensor $\epsilon_{ijkl}$. For the Sp case the $\epsilon$-tensor can
be built from the invariant symplectic tensor
$\omega_{ij}$. The space of antisymmetric bivectors
$\A =\{ a^{[ij]} \}$ is six dimensional. The $\epsilon$-tensor
defines the scalar product on $\A$. This gives a homomorphism
from SL(4) to SO(6). Discreteness of its kernel and equality of the
dimensions of the groups
shows that it is a local isomorphism.

For Sp(2) the space $\A$ is reducible. It decomposes into the space
of tensors proportional to $\omega$ and the space $\A_0$ of
tensors with zero trace, $\A_0 = \{ a^{[ij]}, \omega_{ji}
a^{[ij]} =0 \} $. Therefore here we have a homomorphism
from Sp(2) to SO(5). Again comparison of dimensions shows
that it is  a local isomorphism.

We now turn to quantum groups. The coordinates $x^i$ of the
space $V$ do not commute. Their commutation relations are
given in both cases by
\be
\hR^{ij}_{kl}x^kx^l = qx^ix^j
\ee
for the corresponding $\hR$-matrices [7].
The SL(4) $\hR$-matrix decomposes into two projectors,
\be
\hR = qP^S-q^{-1}P^A ,
\ee
while for Sp(2) we have
\be
\hR = qP^S-q^{-1}P^{A_0} -q^{-5}P^\omega .
\ee
Here $P^S $ projects on symmetric tensors, $P^A$ ($P^{A_0}$)
projects on antisymmetric tensors for SL$_q$(4) (antisymmetric
traceless tensors for Sp$_q$(2)), and $P^\omega$ projects on
the $q$--symplectic tensor.

These decompositions immediately
provide the 6-- and 5--dimensional spaces
needed for constructions of the
isomorphisms. For SL$_q$(4) we consider the
6--dimensional space $\A$ consisting
of tensors $a^{[ij]}$ satisfying
\be P^A a = a,\ee
while for Sp$_q$(2) we have the 5--dimensional
space $\A_0$ of tensors satisfying
\be P^{A_0} a = a.\ee

On the classical level the components
$a^{[ij]}$ commute and isomorphisms
are established by considering transformation properties of
$a^{[ij]}$. The novel feature of quantum spaces
is the noncommutativity
of $a^{[ij]}$. To complete the description of
the spaces $\A$ and $\A_0$
we must fix commutation relations between $a^{[ij]}$.
Moreover, we demand
that these relations are given in terms of some $\hR$--matrix.

In finding such $\hR$--matrices we follow the strategy
of [1--5]. Namely,
we consider $a^{[ij]}$ as composite objects built from
several copies of
the original quantum space. Let $x^i$ and
$\tilde{x}^i$ be two copies of the
original $q$--space. In order to ensure covariance and
consistency we
demand that the commutation relations between $x$ and
$\tilde{x}$ have the form
\be \tilde{x}^i
 x^j = \gamma (\hR^{\pm 1})^{ij}_{kl} x^k \tilde{x}^l,
 \label{e7} \ee
where $\gamma$ is some constant.

The relations for coordinates $x^i$ of a
single $q$--space are described by
projectors comprising an $\hR$--matrix. In
contrast, relations (\ref{e7})
make use of the entire $\hR$--matrix.

We now build a tensor $t^{ij}$ out of two copies $x$ and $y$ of the
original $q$--space,
\be t^{ij} = x^i y^j. \ee
According to the above arguments, an $\hR$--matrix for $t^{ij}$ is
specified by commutation relations between $t^{ij}$ and another copy
$\tilde{t}^{ij}$. Therefore, we have to introduce two more copies
$\tilde{x}^i$ and $\tilde{y}^i$ of the original $q$--space and to
define $\tilde{t}^{ij}=\tilde{x}^i\tilde{y}^j$. Fixing the relations
between the tilded and untilded copies of the original $q$-spaces,
we find relations between $t$ and $\tilde{t}$.

For the quantum Minkowski space [1--5] this procedure resulted in two
$\hR$--matrices. They have the form
\be
\hR_{23}\hR_{12}\hR_{34}\hR_{23}^{-1} \ {\rm and}\
           \hR_{23}\hR_{12}^{-1}\hR_{34}\hR_{23}^{-1}.
\label{e9}
\ee
Here the standard tensor notations are used. For instance,
$\hR_{23}$ is the operator
\be
(\hR_{23})^{ijkl}_{abcd}=\delta^i_a \hR^{jk}_{bc} \delta^l_d
\ee
acting in $V\otimes V\otimes V\otimes V$.

Our aim is to find the closed relations for
the antisymmetric part $a^{[ij]}$ of the tensor $t^{ij}$.
For choices (\ref{e9}) the commutation relations between $a$ and
$\tilde{a}$ are not closed. They include other projections of
the tensors $t$ and $\tilde{t}$. This follows from the fact that
the matrices (\ref{e9}) do not commute with $P^A_{12} P^A_{34}$.
However there is another $\hR$--matrix which commutes with
$P_{12}P_{34}$ for any $P$ of $\hR$. It has the form
\be
\cR_{12\; 34} =
           \hR_{23}\hR_{12}\hR_{34}\hR_{23}.
\label{e11}
\ee
We denote it by $\cR$ to avoid
confusion with the original $\hR$--matrix.

\noindent{\bf Lemma 1.}
$\aR{12}{34}$ satisfies the Yang--Baxter relation
\begin{equation}
     \aR{12}{34}\aR{34}{56}\aR{12}{34}=
     \aR{34}{56}\aR{12}{34}\aR{34}{56}.
\end{equation}
\noindent{\bf Lemma 2.} For and projector $P$ of $\hR$,
\begin{eqnarray}
      P_{12}\aR{12}{34} &=& \aR{12}{34} P_{34}, \nonumber \\
      P_{34}\aR{12}{34} &=& \aR{12}{34} P_{12}.
   \label{e13}
\end{eqnarray}
\noindent{\it Proof of Lemma 1.}  Using the Yang--Baxter relation
for $\hR$ we obtain
\begin{eqnarray}
  \hR_{34}\hR_{23}\aR{34}{56}\hR_{23}\hR_{34} &=&
      \hR_{34}\hR_{23}\hR_{45}\hR_{34}\hR_{56}\hR_{45}\hR_{23}
     \hR_{34} \nonumber \\
  &=& \hR_{34}\hR_{45}(\hR_{23}\hR_{34}\hR_{23}) \hR_{56}\hR_{45}
    \hR_{34} \nonumber \\
  &=& \hR_{34}\hR_{45}\hR_{34}\hR_{23}\hR_{34}
  \hR_{56}\hR_{45}\hR_{34}
  \nonumber \\
 &=& (\hR_{34}\hR_{45}\hR_{34}) \hR_{23}\hR_{56} (\hR_{34}\hR_{45}
     \hR_{34}) \nonumber \\
 &=& \hR_{45}\hR_{34}\hR_{45}\hR_{23}\hR_{56}\hR_{45}\hR_{34}
      \hR_{45} \nonumber \\
 &=& \hR_{45}\hR_{34}\hR_{23} (\hR_{45}\hR_{56}\hR_{45})
       \hR_{34}\hR_{45} \nonumber \\
 &=& \hR_{45}\hR_{34}\hR_{23}\hR_{56}\hR_{45}\hR_{56}\hR_{34}\hR_{45}
 \nonumber \\
 &=& \hR_{45}\hR_{56}\aR{23}{45} \hR_{56}\hR_{45}.
\end{eqnarray}
Therefore,
\begin{eqnarray}
 \aR{12}{34}\aR{34}{56}\aR{12}{34} &=&
     \hR_{23}\hR_{12} (\hR_{34}\hR_{23}\aR{34}{56}\hR_{23}\hR_{34})
       \hR_{12}\hR_{23} \nonumber \\
  &=& \hR_{23}\hR_{12}\hR_{45}\hR_{56} \aR{23}{45}
        \hR_{56}\hR_{45}\hR_{12}\hR_{23} \nonumber \\
  &=& \hR_{45}\hR_{56} (\hR_{23}\hR_{12}\aR{23}{45}\hR_{12}\hR_{23})
       \hR_{56}\hR_{45} \nonumber \\
  &=& \hR_{45}\hR_{56}\hR_{34}\hR_{45} \aR{12}{34}
         \hR_{45}\hR_{34}\hR_{56}\hR_{45} \nonumber \\
  &=& \aR{34}{56} \aR{12}{34} \aR{34}{56},
\end{eqnarray}
which is the Yang--Baxter equation for $\cR$. \\[0.1in]
{\it Proof of Lemma 2.} The Yang--Baxter equation for $\hR$
implies
\be P_{12}\hR_{23}\hR_{12} = \hR_{23} \hR_{12} P_{23} \ee
for any projector $P$ of $\hR$. Thus,
\begin{eqnarray}
  P_{34}\aR{12}{34} &=& (P_{34}\hR_{23}\hR_{34}) \hR_{12}\hR_{23}
   = \hR_{23}\hR_{34}(P_{23}\hR_{12}\hR_{23}) \nonumber \\
   &=& \hR_{23}\hR_{34}\hR_{12}\hR_{23}P_{12}
   = \aR{12}{34} P_{12},
\end{eqnarray}
as stated. The other relation is similarly derived.\\[0.1in]
{\bf Corollary 1.}
\be P_{12}P_{34} \aR{12}{34} = \aR{12}{34} P_{12}P_{34} . \ee
{\bf Corollary 2.}
\be \aR{12}{34}^P = P_{12}P_{34}\aR{12}{34} \ee
satisfies the Yang--Baxter equation. \\[0.1in]
{\em Proof of Corollary 1.} This is an immediate consequence of
 (\ref{e13}). \\[0.1in]
{\em Proof of Corollary 2.} Multiplying the Yang--Baxter equation
for $\aR{12}{34}$ by $P_{12}P_{34}P_{56}$ from either the left
or right and using Lemma 2 we immediately obtain the Yang--Baxter
equation for $\aR{12}{34}^P$.

To produce the $\hR$--matrix (11)
as a result of commutation relations
between bivectors we choose the commutation relations between
 untilded and tilded quantities to be
\be
x^1 \tilde{x}^2=\hR_{12} \tilde{x}^1 x^2, \;\;\;
x^1 \tilde{y}^2=\hR_{12} \tilde{y}^1 x^2, \;\;\;
y^1 \tilde{x}^2=\hR_{12} \tilde{x}^1 y^2, \;\;\;
y^1 \tilde{y}^2=\hR_{12} \tilde{y}^1 y^2.
\ee
In general, one can add four arbitrary numerical factors on the
RHSs of (20). However,
they are not essential for the purposes of the
present paper, and we will not introduce them in what follows.
Relations (20) lead to the relations
\be t^{12} \tilde{t}^{34} = \aR{12}{34} \tilde{t}^{12} t^{34}, \ee
which are consistent by Lemma 1. By Corollary 1, the objects
\begin{equation}
    t_P^{12} = P_{12} x^1 y^2, \quad
    \tilde{t}_P^{12} = P_{12} \tilde{x}^1 \tilde{y}^2,
\end{equation}
for any $P$ of $\hR$, satisfy the commutation relations
\be t^{12}_P \tilde{t}^{34}_P = \aR{12}{34}^P
 \tilde{t}_P^{12} t_P^{34}. \ee
These relations are seen to be consistent by Corollary 2.

Under the group action we have the transformation laws
\begin{equation}
     x \rightarrow T x, \quad
     y \rightarrow T y, \quad
     \tilde{x} \rightarrow T \tilde{x}, \quad
     \tilde{y} \rightarrow T \tilde{y}.
\end{equation}
The quantum matrix elements $T^i_j$ satisfy the relations
\begin{equation}
        \hR_{12} T_1 T_2 = T_1 T_2 \hR_{12}.
\end{equation}
For any
$P$ a polynomial in $\hR$, the tensor product objects
transform as
\begin{equation}
    t^P_{12} \rightarrow  T_1 T_2 P_{12} t^P_{12}
         = T^P_{12} t^P_{12}.
\end{equation}
Here we have defined $T^P_{12}= T_1 T_2 P_{12}$, a tensor product
quantum matrix. \\[0.1in]
{\bf Proposition.} The elements of $T_{12}^P$ satisfy the
commutation relations
\begin{equation}
       \aR{12}{34}^P T^P_{12} T^P_{34} =
       T^P_{12} T^P_{34}\aR{12}{34}^P
\end{equation}
{\em Proof. }Follows from the commutation relations for
elements of $T$.

Therefore, to demonstrate the local isomorphisms we need only to
show the coincidence of the corresponding $\hR$--matrices.

In this section we have seen
that the consideration of $t$ as a secondary object built
from $x$ and $y$ leads to several different $\hR$--matrices.
Direct inspection shows that $\cR$ is the unique $\hR$--matrix
(up to taking the inverse) satisfying (\ref{e13}).
We do not give more precise
statements about the possible choices of $\cR$--matrices since our
aim here is only to establish the isomorphisms of low dimensional
quantum groups.

\newpage
\noindent {\Large \bf 3. SL$_q$(4)$\sim$SO$_{q}$(6). }

In this section we investigate the $\cR$--matrix
\begin{equation}
   \R{12}{34} = \hR_{23}\hR_{34}\hR_{12}\hR_{23}P^A_{12}P^A_{34}
   \label{e21}
\end{equation}
for the
SL(n) case in some detail, and display explicitly the
isomorphism between SO$_{q}$(6) and SL$_{q}$(4). Eq. (\ref{e21})
corresponds to (19) for $P=P^A$ and gives the commutation
relations between $a$ and $\tilde{a}$.

\noindent {\bf Proposition.}
$\R{12}{34}$ satisfies the cubic equation
\def\AA{_{[\, ][\, ]}}
\def\K#1#2{{\cal K}_{[#1][#2]}}
\def\I#1#2{{\Pi}_{[#1][#2]}}
\begin{equation}
    (\hBR\AA-q^{2}{\Pi }\AA)
    (\hBR\AA+{\Pi}\AA)
    (\hBR\AA-q^{-4}{\Pi}\AA) = 0,
\end{equation}
where
\begin{equation}
  \I{12}{34} = P^A_{12} P^A_{34}.
\end{equation}
{\em Proof.} Eq. (18) implies that
\be
   \R{12}{34} = \I{12}{34}\R{12}{34}.
\ee
One may verify that
\be
   \R{12}{34}^2 = \lambda^2 q^{-1}(q+q^{-1}) \R{12}{34}
    +\lambda q^{-2} (q+q^{-1})^2 \K{12}{34} + \I{12}{34},
   \label{e28}
\ee
where
\begin{equation}
  \K{12}{34} = P^A_{12} P^A_{34} \hR_{23} P^A_{12} P^A_{34}.
\end{equation}
Next, we find
\begin{equation}
   \R{12}{34}\K{12}{34}=\lambda\R{12}{34}+q^{-2}\K{12}{34}.
\end{equation}
Multiplying eq. (\ref{e28})
for $\hBR^2\AA$ once again by $\hBR\AA$ we
get an expression for $\hBR^3\AA$ which contains $\hBR\AA$,
 ${\cal K}\AA$ and ${\Pi}\AA$ linearly.
The cubic equation involving only
$\hBR$ follows immediately. The proof is finished.

If some combination of $\hBR\AA$ and
${\cal K}\AA$ is zero or ${\Pi}\AA$, then $\hBR\AA$
satisfies a characteristic
equation of lower degree. However, formulas
(39) below for the traces of the projectors show that this does not
happen  when $n>3$.

{}From the characteristic equation for $\hBR\AA$ we find
the projector decomposition
\begin{eqnarray}
   \R{12}{34} &=& q^{ 2}\P{12}{34}^S -\P{12}{34}^A
       + q^{-4}\P{12}{34}^T, \nonumber \\
   \I{12}{34} &=& \P{12}{34}^S +\P{12}{34}^A
       + \P{12}{34}^T,
\end{eqnarray}
where
\begin{eqnarray}
    \P{12}{34}^S &=& {(\R{12}{34}+\I{12}{34})
      (\R{12}{34}-q^{-4}\I{12}{34}) \over
      q^{4} (1+q^{-2})(1-q^{-6}) },\nonumber \\
    \P{12}{34}^A &=& {(\R{12}{34}-q^{ 2}\I{12}{34})
      (\R{12}{34}-q^{-4}\I{12}{34}) \over
       (1+q^{2})(1+q^{-4}) },\nonumber \\
    \P{12}{34}^T &=& {(\R{12}{34}-q^{ 2}\I{12}{34})
      (\R{12}{34}+\I{12}{34}) \over
       (q^{-4}-q^{ 2})(q^{-4}+1) }.
\end{eqnarray}

Let us compute the dimensionality of these projectors. Using the
cyclic property of trace, and
properties of the SL$_q$(n) $\hR$--matrix,
we obtain
\def\Tr{{\rm Tr}}
\begin{eqnarray}
    \Tr\R{12}{34} &=& \Tr P^A_{12}P^A_{34}\hR_{23} \hR_{12}
      \hR_{34} \hR_{23} \nonumber \\
     & = & {1\over q+q^{-1}}\Tr\[q\hR_{12}\hR_{34}
            -(q-q^{-1})\hR_{12}\hR_{23}
             - q^{-2} \Tr \hR_{12} \hR_{34} \hR_{23}\] \nonumber \\
    & =&  {n(n-1)\over 24 q^4} \[
            2q^6 n(n+1)-3 q^4(n+1)(n-2)+(n-2)(n-3) \].
\end{eqnarray}
Similarly,
\begin{eqnarray}
   \Tr \K{12}{34} &=
   &\Tr P^A_{12} P^A_{23} \hR_{23} P^A_{12} P^A_{34}
\nonumber \\ &=&
 {1\over (q+q^{-1})^2}
     \Tr\[ q^2 \hR_{23}-
     2q \hR_{12}\hR_{23}+\hR_{12}\hR_{34}\hR_{23} \]
   \nonumber \\
    & =&  {n(n-1)\over 24 q(q^2+1)} \[
            q^4 (5n-6)(n+1)-4 q^2 n(n-2)-(n-2)(n-3) \].
\end{eqnarray}
We then find,
\begin{eqnarray}
    \Tr \P{12}{34}^S &=& {1\over12} n^2(n+1)(n-1) ,\nonumber \\
    \Tr \P{12}{34}^A &=& {1\over8} n (n+1)(n-1)(n-2), \nonumber \\
    \Tr \P{12}{34}^T &=& {1\over 24} n(n-1)(n-2)(n-3).
\end{eqnarray}
The projector decomposition
of $\hBR\AA$ has the following Young tableaux representation:

\begin{picture}(150,50)(-130,0)
\thicklines
\put(10,10){\framebox(10,10){}}
\put(10,20){\framebox(10,10){}}
\put(25,17){$\times$}\put(25, 37){$\hBR$}
\put(40,10){\framebox(10,10){}}
\put(40,20){\framebox(10,10){}}
\put(55,17){=}
\put(70,10){\framebox(10,10){}}
\put(70,19.9){\framebox(10,10){}}
\put(79.9,10){\framebox(10,10){}}
\put(79.9,19.9){\framebox(10,10){}}
\put(95,17){+}\put(74, 37){${\cal P}^S$}
\put(110,10){\framebox(10,10){}}
\put(110,19.9){\framebox(10,10){}}
\put(110,0.1){\framebox(10,10){}}
\put(119.9,19.9){\framebox(10,10){}}
\put(135,17){+}\put(114, 37){${\cal P}^A$}
\put(150,10){\framebox(10,10){}}
\put(150,20){\framebox(10,10){}}
\put(150, 0){\framebox(10,10){}}
\put(150,-10){\framebox(10,10){}}
\put(148, 37){${\cal P}^T$}
\end{picture}
\begin{equation}\end{equation}

For $n=4$,  ${\cal P}^S\AA$, ${\cal P}^A\AA$ and
${\cal P}^T\AA$ have the dimensionality 20, 15 and 1, respectively.
These numbers correspond exactly to the traces of the symmetric,
antisymmetric and singlet projectors, respectively, of the
standard SO$_q$(6) $\hR$--matrix [7]. Moreover,
on the space of antisymmetric tensors, the projector $\Pi$ takes
the eigenvalue 1, and for $n=4$
eq. (29) is the same as the usual cubic equation
for $q\hR_{{\rm SO}_q(6)}$.
Now we show that not only the characteristic
equation but the $\hR$--matrices themselves are the same.

\noindent {\bf Proposition.} For $n=4$, the matrix
$q^{-1}\R{12}{34}$ coincides, up to a change of basis, with
the standard
SO$_q$(6) $\hR$--matrix.\\[0.1in]
{\em Proof.} We explicitly find a basis in which the $\hR$--matrices
coincide. The basis is
\begin{eqnarray}
   a^1 &=&   x^1 y^2 - q x^2 y^1,\nonumber \\
   a^2 &=&   x^1 y^3 - q x^3 y^1,\nonumber \\
   a^3 &=&   x^1 y^4 - q x^4 y^1,\nonumber \\
   a^4 &=&   x^2 y^3 - q x^3 y^2,\nonumber \\
 - a^5 &=&   x^2 y^4 - q x^4 y^2,\nonumber \\
   a^6 &=&   x^3 y^4 - q x^4 y^3.
\end{eqnarray}
These are proportional to the six objects $P_{12}^A x^1 y^2$.
(In fact, $(P^A)_{kl}^{ji} x^k y^l$ for $i<j$ are proportional by
the same amount to $x^i y^j -  q x^j y^i$.
With hindsight, we have
inserted a minus sign in the definition of $a^5$.)
We similarly define $\tilde{a}^1$ through $\tilde{a}^6$
by putting tildes everywhere. Then, one obtains the following
commutation relations:
\def\ta{\tilde{a}}
\begin{eqnarray}
    {a}^i \ta^i &=& q^{2} \ta^i{a}^i, \nonumber \\
    {a}^i \ta^k &=& q \ta^k{a}^i,
       \nonumber \\
    {a}^k \ta^i &=& q \ta^i{a}^k+(q^2-1)
        \ta^k{a}^i, \nonumber\\
    {a}^6 \ta^1 &=&  \ta^1 {a}^6 , \nonumber \\
    {a}^5 \ta^2 &=&  \ta^2 {a}^5-q^{-1}(q^2-1)
        \ta^1 {a}^6  , \nonumber \\
    {a}^4 \ta^3 &=&  \ta^3 {a}^4-q^{-1}(q^2-1)
        \ta^2 {a}^5 -q^{-2}(q^2-1)\ta^1{a}^6  , \nonumber \\
    {a}^3 \ta^4 &=&  \ta^4 {a}^3-q^{-1}(q^2-1)
        \ta^2 {a}^5 -q^{-2}(q^2-1)\ta^1{a}^6  , \nonumber \\
    {a}^2 \ta^5 &=&  \ta^5 {a}^2-q^{-1}(q^2-1)
        \ta^3 {a}^4 -q^{-1}(q^2-1)\ta^4{a}^3\nonumber \\& &
        +q^{-2}(q^2-1)^2 \ta^2{a}^5 - q^{-3}(q^2-1)
            \ta^1{a}^6 , \nonumber \\
    {a}^1 \ta^6 &=&  \ta^6 {a}^1-q^{-2}(q^2-1)
        \ta^3 {a}^4 -q^{-2}(q^2-1)\ta^4{a}^3\nonumber \\
        & &    -q^{-3}(q^2-1) \ta^2{a}^5
        -q^{-1}(q^2-1) \ta^5{a}^2
        +q^{-4}(q^6-q^4-q^2+1) \ta^1{a}^6.
\end{eqnarray}
In these expressions, $i>k$
and $i\neq 7-k$. A direct comparison shows
that relations (42) coincide with the
relations
\be  {a}^i \ta^j = q(\hR_{{\rm SO}_{q}(6)})^{ij}_{kl}
     \ta^k {a}^l  \ee
for the SO$_{q}$(6) $\hR$--matrix [7]
\begin{eqnarray}
  \hR^{ji}_{kl} &=& \delta^j_l\delta^i_k (1+\delta^{i-j}(q-1)
      +\delta^{i+j-7}(q^{-1}-1))+\lambda \theta(l-k)\delta^j_k
       \delta^i_l \nonumber \\ & & -\lambda\theta(i-k)
       q^{\rho_i-\rho_k} \delta_{k+l-7}\delta^{i+j-7},
\end{eqnarray}
where $\lambda=q-q^{-1}$,  $\rho_{i}=3-i$ for $i<4$ otherwise
$\rho_{i}=4-i$ and $\theta(i)=1$ if $i>0$ otherwise $\theta(i)=0$.
 The proof is finished.

According to (27) this coincidence gives the
isomorphism of the corresponding quantum groups.

To conclude this section we discuss the following question. The
$\cR$--matrix (\ref{e21}) was obtained by moving $a$ through
$\tilde{a}$. For this
we need to know neither the commutation relations
between $x$ and $y$ nor those between $\tilde{x}$ and $\tilde{y}$.
Thus we have two different ways of defining relations between the
components of $a$. The
first is to find the consistent relations between
$x$ and $y$. The second way is to use the antisymmetric
projector of the $\cR$--matrix (\ref{e21}). In general, these
two ways lead to different results. However, in the case studied
they coincide for any choice (\ref{e7}) of the $xy$ commutation
relations. This amounts to the statement that the antisymmetric
projector $\P{12}{34}^A$ vanishes on $a^{[12]}a^{[34]}$.
A direct calculation shows that even the factor
$(\R{12}{34}-q^{2}\I{12}{34})$ vanishes on
$a^{[12]} a^{[34]}= P^A_{12} P^A_{34} x^1 y^2 x^3 y^4$, which is
a stronger statement. The reason is that the expression
$a^{[12]}=P^A_{12}x^1 y^2$
gives only degenerate antisymmetric tensors
satisfying extra relations. Such a degeneracy occurs also in the
Lorentz case. In [1,3] it was removed by introducing
two extra
copies $\xi,\eta$ of spinors with consistent commutation relations
between all the copies of the $q$--spinors. We could do the
same in our case. However, knowing the entire
$\hR$--matrix one can simplify this procedure using the following
lemma.\\[0.1in]
{\bf Lemma.}  Assume that commutation relations for the coordinates
$x^i$ of some quantum space are given by a single projector $P$ of
an $\hR$--matrix,
\be P_{12} x^1 x^2 = 0 . \label{exx} \ee
Let $y$ be another copy of the same quantum space, $P_{12}y^1 y^2=0$.
Suppose that the $xy$ relations are given by
\be y^1 x^2 = -\alpha^{\mp 1} \hR_{12}^{\pm 1} x^1 y^2, \ee
where $\alpha$ is the coefficient with which the projector $P$
enters the $\hR$--matrix, $\hR=\alpha P+\ldots$. Then  $x+y$ is
again a quantum vector, $P_{12}(x+y)^1 (x+y)^2=0$.

The proof is straightforward. Since the conditions of the lemma
are  satisfied for the $q$--orthogonal spaces, we remove
the degeneracy by considering sums $a+a'+a''+\ldots$ with
appropriate commutation relations between the copies of $a$.
On such sums only the whole projector $\P{12}{34}^A$ vanishes.

\vskip 0.65cm
\noindent{\Large\bf 4. Sp$_q$(2)$\sim$SO$_{q^2}$(5).}\\[0.1cm]

The $\hR$-matrix for Sp$_q$(n) contains three projectors. The tensor
$t^{ij}$ decomposes now into three components, $t^{ij}=a^{[ij]}+\tau
\omega^{ij}+s^{(ij)}$. Here $\omega^{ij}$ is the $q$--symplectic
tensor. The component $a^{[ij]}$ is traceless, $\omega_{ji}
a^{[ij]} =0$, and the component $s^{(ij)}$ is $q$--symmetric.

The component $a$ is now singled out by the projector $P^{A_0}$.
The matrix $
   \R{12}{34} =
   \hR_{23}\hR_{34}\hR_{12}\hR_{23}P^{A_0}_{12}P^{A_0}_{34}$
no longer satisfies a cubic characteristic
equation for general $n$. This is to be expected since
the rules of multiplication of Young tableaux are more
complicated for symplectic spaces.
It is straightforward to
find the characteristic equation for $\cR\AA$ and then its
projector decomposition.
Here we discuss only the case Sp$_q$(2). Then, $\R{12}{34}$
does satisfy a cubic characteristic equation which
in addition coincides with
the characteristic equation for $\hR_{{\rm SO}_{q^2}(5)}$.
This suggests the possibility of the local
isomorphism  Sp$_q$(2)$\sim$SO$_{q^2}$(5), which we briefly
demonstrate below.

Let
\begin{eqnarray}
a^{1}&=& x^{1}y^{2}-qx^{2}y^{1}, \nonumber \\
a^{2}&=& x^{1}y^{3}-qx^{3}y^{1}, \nonumber \\
a^{3}&=&(x^1 y^4-x^4 y^1-q^{-1}x^2 y^3+q x^3 y^2)/\sqrt{1+q^{-2}}
           , \nonumber \\
a^{4}&=&x^{2}y^{4}-qx^{4}y^{2}, \nonumber \\
-a^{5}&=&x^{3}y^{4}-qx^{4}y^{3}.
\end{eqnarray}
These are proportional to the five $a^{[ij]}$. We define tilded
quantities in the same manner.
An explicit calculation results
in the following $a\tilde{a}$ relations:
\begin{eqnarray}
{a}^i \ta^i &=& q^2 \ta^i{a}^i ,\qquad
        \quad\qquad\qquad \qquad i\neq 3,\nonumber \\
{a}^i \ta^j &=& \ta^j{a}^i+(q^2-q^{-2})\ta^i{a}^j,
  \;\;\qquad i<j,\; i+j\neq 6 , \nonumber \\
{a}^i \ta^j &=& \ta^j{a}^i ,\qquad\qquad\qquad\qquad\qquad
     i>j,\; i+j\neq 6,\nonumber\\
{a}^{5} \ta^1 &=&q^{-2} \ta^1{a}^{5},\nonumber \\
{a}^{4} \ta^2 &=&q^{-2}\ta^2{a}^{4}-q^{-2}(q^2-q^{-2})\ta^1{a}^{5},
    \nonumber \\
{a}^{3} \ta^3 &=& \ta^3{a}^{3}-
 q^{-3}(q^2-q^{-2})\ta^1{a}^{5} -
q^{-1}(q^2-q^{-2})\ta^2{a}^{4}, \nonumber \\
{a}^{2} \ta^{4} &=& q^{-2} \ta^{4}{a}^{2}+
(q^2-q^{-2})(1-q^{-2})\ta^2{a}^{4}-
q^{-4}(q^2-q^{-2})\ta^1{a}^{5} \nonumber \\ & &
-q^{-1}(q^2-q^{-2})\ta^3{a}^{3}, \nonumber \\
{a}^{1}\ta^{5} &=& q^{-2} \ta^{5}{a}^{1}+
(q^2-q^{-2})(1-q^{-6})\ta^1{a}^{5}-
q^{-4}(q^2-q^{-2})\ta^2{a}^{4} \nonumber \\ & & -
q^{-2}(q^2-q^{-2})\ta^4{a}^{2} -
q^{-3}(q^2-q^{-2})\ta^3{a}^{3}.
\end{eqnarray}
As in the SL(4) case the
direct inspection shows that these relations
coincide with
\be
{a}^i\ta^j =  (\hR_{SO_{q^2}(5)})^{ij}_{kl}\ta^k{a}^l,
\ee
for the SO$_{q^2}$(5) $\hR$--matrix [7],
\begin{eqnarray}
  \hR^{ji}_{kl} &=& \delta^j_l\delta^i_k (1+\delta^{i-j}(q^2-1)
      +\delta^{i+j-6}(q^{-2}-1)-\delta^{i-3}\delta^{j-3}\lambda^2)
  \nonumber \\ & &
+(q^2-q^{-2})\theta(l-k)\delta^j_k
       \delta^i_l
 -(q^2-q^{-2})\theta(i-k)
       q^{2\rho_i-2\rho_k} \delta_{k+l-6}\delta^{i+j-6},
\end{eqnarray}
where $\rho_1,\ldots,\rho_5= 3/2,1/2,0,-1/2,-3/2$. Once again, the
use of (27) establishes the needed local isomorphism.


\vskip 0.25cm
\noindent{\bf Acknowledgements.} \\
We would like to thank H. Ewen, Michael Schlieker,
W. B. Schmidke, J. Wess and B. Zumino for valuable discussions.

\end{document}